\documentclass[aps,pra,10pt,amsmath,twocolumn,floatfix,superscriptaddress,longbibliography,eqsecnum]{revtex4-1}
\usepackage[dvipsnames]{xcolor}
\usepackage[colorlinks=true,bookmarks=false,linkcolor=NavyBlue,urlcolor=NavyBlue,citecolor=NavyBlue,breaklinks]{hyperref}
\usepackage{amsmath,amsfonts,amssymb,amsthm}
\usepackage{mathtools}
\usepackage{tabularx}
\usepackage{graphicx}
\usepackage{physics}
\usepackage{newtxtext,inconsolata}
\usepackage{enumitem}
\usepackage{setspace}

\theoremstyle{plain}

\theoremstyle{definition}

\theoremstyle{remark}

\newcommand{\intall}{\int_{-\infty}^{\infty}}

\newcommand{\ceil}[1]{\left\lceil#1\right\rceil}

\newcommand{\bk}[1]{\qty(#1)}
\newcommand{\Bk}[1]{\qty[#1]}
\newcommand{\BK}[1]{\qty{#1}}

\newcommand{\mc}[1]{\mathcal #1}
\newcommand{\mb}[1]{\mathbb #1}

\newcommand{\SigmaX}{\Gamma}
\newcommand{\SigmaZ}{\Upsilon}
\newcommand{\SigmaW}{R}

\DeclareMathOperator*{\argmax}{arg\,max}

\DeclareMathOperator{\expect}{\mathbb E}

\DeclareMathOperator{\MSE}{MSE}

\newcommand{\fig}[3]{\begin{figure*}[htbp!]\centerline{\includegraphics[width=#1\textwidth]{#2}}
\caption{\label{#2}#3}
\end{figure*}}
\begin{document}

\title{Spectrum analysis with quantum dynamical systems. II. Finite-time analysis}

\author{Xinyi Sui}
\affiliation{Department of Physics, National University of Singapore,
2 Science Drive 3, Singapore 117551}
\author{Mankei Tsang}
\email{mankei@nus.edu.sg}
\homepage{https://blog.nus.edu.sg/mankei/}
\affiliation{Department of Electrical and Computer Engineering,
  National University of Singapore, 4 Engineering Drive 3, Singapore
  117583}

\affiliation{Department of Physics, National University of Singapore,
  2 Science Drive 3, Singapore 117551}

\date{\today}


\begin{abstract}
  The prequel to this work [Ng \emph{et al.}, Phys.~Rev.~A
  \textbf{93}, 042121 (2016)] proposes the method of spectral photon
  counting to enhance noise spectroscopy with an optical
  interferometer. While the predicted enhancement over homodyne
  detection is promising, the results there are derived by taking an
  asymptotic limit of infinite observation time; their validity for a
  finite time remains unclear.  To validate the theory, here we
  perform a numerical study of a finite-time model. Assuming that the
  signal is an Ornstein--Uhlenbeck process with an unknown variance
  parameter, we evaluate the Fisher information for homodyne detection,
  a lower bound on the Fisher information for spectral photon counting,
  and a quantum upper bound, all without taking the infinite-time
  limit. To confirm that the Fisher-information quantities are
  satisfactory precision measures, we also compute the errors of the
  maximum-likelihood estimator by Monte-Carlo simulations. The results
  demonstrate that the Fisher-information quantities and the
  estimation errors all smoothly approach their asymptotic limits, and
  the advantage of spectral photon counting over homodyne detection
  can remain substantial for finite times.
\end{abstract}

\maketitle
\section{Introduction}
Noise spectroscopy is the measurement of the power spectral density of
a stationary stochastic signal. It is a fundamental task for many
applications of optical interferometry, including thermometry
\cite{dedyulin22}, observation of the stochastic gravitational-wave
background \cite{christensen18}, test of wavefunction-collapse models
\cite{nimmrichter}, and test of novel quantum gravity theories
\cite{mcculler22,vermeulen25}.  The prequel to this work \cite{ng16},
hereafter called Paper I, proposes the method of spectral photon
counting (SPC) to enhance noise spectroscopy with an optical
interferometer. This method is nothing but an optical spectrometer in
principle: it first separates the output light into different channels
in terms of the spectral modes and then counts the photons in each
spectral mode. Paper I finds that, surprisingly, the Fisher
information for SPC with a coherent optical state is far superior to
that of homodyne detection in a weak-signal regime. SPC is also
quantum-optimal, in the sense that its Fisher information coincides
with a quantum upper bound for any measurement. The idea has since
been adopted by the Gravity from the Quantum Entanglement of Space
Time (GQuEST) project to test certain quantum gravity theories
\cite{mcculler22,vermeulen25}. Generalizations of SPC for nonclassical
states have also been proposed
\cite{gorecki22,shi23,noise_spec_pra,gardner25}.

Two open questions remain regarding Paper I. The first is that the
Fisher-information expressions there are all derived under the
approximation of stationary process and long observation time (SPLOT)
and their reliability for any finite time is unknown.  The second
question is whether the Cram\'er-Rao bound, given by the inverse of
the Fisher information, is tight to the estimation error for a finite
time in noise spectroscopy. The second question is especially
concerning, since the Cram\'er-Rao bound is guaranteed to be
achievable only in an asymptotic long-time limit
\cite{shumway_stoffer} and may be unreliable for a finite sample
size---it may be too loose for some problems \cite{bell,qzzb} or may
be severely violated by biased estimators for some other problems
\cite{vaart,minimax_jmo}. The superiority of SPC over homodyne
detection in reality is therefore questionable.

This paper addresses the two open questions by performing a numerical
study of a finite-time model.  Assuming that the signal is an
Ornstein--Uhlenbeck process with an unknown variance parameter, we
compute the Fisher information for homodyne detection, a lower bound
on the Fisher information for SPC, and a quantum upper bound, all for
finite observation times.  We also evaluate the finite-time errors of
SPLOT-based maximum-likelihood estimators via Monte-Carlo
simulations. The results show that the finite-time results all
approach their asymptotic limits smoothly as the observation time
increases, demonstrating no major discrepancy with the SPLOT-based
results.  Our numerical study hence verifies that our SPLOT-based
results are adequate benchmarks and SPC can remain superior to
homodyne detection for finite times.


\section{\label{sec_theory}Theory}

\subsection{\label{sec_hom}Homodyne detection}
We first present the background theory used in our numerical analysis.
Define discrete time values in terms of a time interval $\delta t > 0$
as
\begin{align}
t_j &\equiv t_0 + j\delta t, & j &= 0,1,\dots,M-1,
&
T &\equiv M \delta t.
\end{align}
$T$ is the total observation time. Let $\{Y(t_j):j = 0,1,\dots,M-1\}$
be an observation process obtained by homodyne detection. Assume that
$Y$ is a zero-mean Gaussian random process with covariance matrix
\begin{align}
\Sigma_{jk}(\theta) &\equiv \expect_\theta\Bk{Y(t_j)Y(t_k)},
\end{align}
where $\theta \in \mb R^p$ is a vectoral parameter and
$\expect_\theta$ denotes the expectation in terms of the probability
density $f_Y^{(\mathrm{hom})}(y|\theta)$ of $Y$ conditioned on
$\theta$. The negative log-likelihood function becomes
\begin{align}
  -\ln f_Y^{(\mathrm{hom})}(y|\theta) &= \frac{1}{2} y^\top \Sigma(\theta)^{-1} y  + 
                                        \frac{1}{2} \ln \det \Sigma(\theta),
\label{loglike_hom}
\end{align}
where $\top$ denotes the transpose and a $\theta$-independent additive
constant has been omitted; such constants will always be omitted
hereafter.  The Fisher information matrix is then given by \cite{kay}
\begin{align}
J_{\mu\nu}^{(\mathrm{hom})}(\theta) &= \trace\Bk{\Sigma(\theta)^{-1} 
\pdv{\Sigma(\theta)}{\theta_\mu}\Sigma(\theta)^{-1} \pdv{\Sigma(\theta)}{\theta_\nu}},
\label{Jhom}
\end{align}
where $\trace$ denotes the trace.

We consider a simplified but representative model of an optical
interferometer: one continuous-wave optical beam under weak phase
modulation. More complicated configurations such as the Michelson
interferometer follow very similar physics
\cite{ng16,noise_spec_pra}. For noise spectroscopy, assume that the
signal $X(t)$ applied by phase modulation is also a zero-mean Gaussian
process. Assume further that the modulation is weak and the mean
optical field has been nulled, such that homodyne detection of the
phase quadrature gives
\begin{align}
Y(t) &= X(t) + Z(t),
\end{align}
where $Z(t)$ is another zero-mean Gaussian process that is independent
from $X(t)$ and models the additive noise in homodyne detection.  The
covariance matrix of $Y$ becomes
\begin{align}
\Sigma(\theta) &= \SigmaX(\theta) + \SigmaZ,
\end{align}
where $\SigmaX(\theta)$ is the parameter-dependent covariance matrix
of $X$ and $\SigmaZ$ is the covariance matrix of $Z$. If the input
beam is in a coherent state, the noise covariance matrix is given by
\begin{align}
\SigmaZ_{jk} &= \frac{1}{4 \mc N \delta t} \delta_{jk},
\label{SigmaZ}
\end{align}
where $\mc N$ is the average photon flux of the optical beam.


If the phase quadrature of the input beam is squeezed and the
squeezing bandwidth is much wider than that of $X$
\cite{walls_milburn}, then $\Upsilon$ is simply reduced by a constant
factor. As this case is a theoretically trivial modification of the
coherent-state case, we do not consider it for brevity and focus on
the coherent-state case.

If $X(t)$ and thus $Y(t)$ are stationary, the covariance is a function
$C_\theta$ of $t_k-t_j$ only, that is,
\begin{align}
  \expect_\theta\Bk{Y(t_j)Y(t_k)} &= \Sigma_{jk}(\theta) = C_\theta(t_k-t_j).
\end{align}
Let the Fourier transform of $Y(t)$ be
\begin{align}
\tilde Y(\omega_m) &\equiv \frac{1}{\sqrt{T}} \sum_{j=0}^{M-1} Y(t_j) \exp(i\omega_m t_j) \delta t,
\\
\omega_m &\equiv \frac{2\pi m}{T}, \quad 
m = \ceil{-\frac{M-1}{2}}, \dots, \ceil{\frac{M-1}{2}},
\label{freq}
\end{align}
where $\ceil{\cdot}$ denotes the ceiling function. The power spectral
density is then defined as
\begin{align}
S_Y(\omega_m|\theta) &\equiv \expect_\theta\Bk{\abs{\tilde Y(\omega_m)}^2}
\label{SY}
\\
&= \sum_{k=0}^{M-1} C_\theta(\tau_k) \exp(i\omega_m \tau_k) \delta t.
\end{align}
In the SPLOT limit, the negative log-likelihood function is given by
the Whittle approximation \cite[Eq.~(4.85)]{shumway_stoffer}
\begin{widetext}
\begin{align}
-\ln f_Y^{(\mathrm{hom,SPLOT})}(y|\theta) 
&=
\frac{1}{2} \Bk{\ln S_Y(0|\theta) + \frac{|\tilde y_0|^2}{S_Y(0|\theta)}}
+ \sum_{m > 0} 
\Bk{\ln S_Y(\omega_m|\theta) + \frac{|\tilde y_m|^2}{S_Y(\omega_m|\theta)}},
\label{loglike_hom_SPLOT}
\\
\tilde y_m &\equiv \frac{1}{\sqrt{T}} \sum_{j=0}^{M-1} y(t_j) \exp(i\omega_m t_j) \delta t.
\label{y_freq}
\end{align}
Note that $\tilde y_0$ is special because it is real, whereas all the
other $\tilde y_m$'s are complex. Note also that, since $y(t)$ is
real, $\tilde y_m = \tilde y_{-m}^*$, and there is no need to consider
$\tilde y_m$ with negative $m$ in the likelihood function. The Fisher
information in the SPLOT limit (together with the $\delta t \to 0$
limit) is \cite{vantrees3}
\begin{align}
J_{\mu\nu}^{(\mathrm{hom,SPLOT})}(\theta) 
&= 
\frac{T}{4\pi} \intall \Bk{\pdv{}{\theta_\mu} \ln S_Y(\omega|\theta)}
\Bk{\pdv{}{\theta_\nu} \ln S_Y(\omega|\theta)} \dd\omega
\\
&= 
\frac{T}{4\pi} \intall 
\frac{1}{\{1+1/[4\mc N S_X(\omega|\theta)]\}^2}\Bk{\pdv{}{\theta_\mu} \ln S_X(\omega|\theta)}
\Bk{\pdv{}{\theta_\nu} \ln S_X(\omega|\theta)} \dd\omega,
\label{Jhom_SPLOT}
\end{align}
\end{widetext}
where $S_X$ is the power spectral density of $X$, defined in the same
way as Eq.~(\ref{SY}). Sec.~\ref{sec_fish} will compare this
expression with the exact Eq.~(\ref{Jhom}) numerically, while
Sec.~\ref{sec_ML} will use Eqs.~(\ref{loglike_hom_SPLOT}) and
(\ref{y_freq}) to compute the maximum-likelihood estimator given
  a time series $\{y(t_j):j = 0,1,\dots,M-1\}$. We use the Whittle
approximation given by Eq.~(\ref{loglike_hom_SPLOT}) instead of the
exact Eq.~(\ref{loglike_hom}) because the former can be evaluated much
more quickly---$\tilde y_m$ can be computed using the fast Fourier
transform while $S_Y(\omega_m|\theta)$ contains much fewer entries
than $\Sigma(\theta)$ and is thus easier to handle. We will see in
Sec.~\ref{sec_ML} that the approximation still gives satisfactory
results.

\subsection{\label{sec_SPC}Spectral photon counting (SPC)}
For the SPC method described in Paper I, the measurement outcome is a
doubly stochastic process. For a coherent-state input beam and a given
signal $X(t)$, each photon count $L_m$ for a spectral mode is an
independent Poisson random variable with conditional mean
\begin{align}
\expect\bk{L_m|X} & = \mc N\abs{\tilde X_m}^2,
\\
\tilde X_m &\equiv \frac{1}{\sqrt{T}}\sum_{j=0}^{M-1} X(t_j) \exp(i\omega_m t_j) \delta t,
\end{align}
where $\omega_m$ is the frequency of the spectral mode given by
Eqs.~(\ref{freq}) relative to the carrier frequency of the optical
beam. $L_m$ is doubly stochastic because $X(t)$ is a random process as
well.  The likelihood function and the Fisher information
$J^{(\mathrm{SPC})}$ become difficult to compute exactly, but we can
use a lower bound due to Stein \emph{et al.}\ \cite{stein14}, given by
\begin{align}
J^{(\mathrm{SPC})}(\theta) &\ge D(\theta)^\top E(\theta)^{-1} D(\theta)
\nonumber\\
&\quad \equiv J^{(\mathrm{SPC,Stein})}(\theta),
\label{stein}
\\
D_{m\mu}(\theta) &\equiv \pdv{}{\theta_\mu} \expect_\theta\bk{L_m},
\\
E_{mn}(\theta) &\equiv \expect_\theta\bk{L_m L_n} - \expect_\theta\bk{L_m}\expect_\theta\bk{L_n}.
\end{align}
The lower bound guarantees that the actual Fisher information must be
above it. The bound is easy to compute because it depends on only the
mean $\expect_\theta(L_m)$ and covariance matrix $E$ of $L_m$.  Let
\begin{align}
\tilde X_m &= \sum_j V_{mj} X(t_j),
\\
V_{mj} &\equiv \frac{\delta t}{\sqrt{T}} \exp(-i\omega_m t_j).
\end{align}
The mean of $L_m$ becomes
\begin{align}
\expect_\theta(L_m)
&= \mc N \expect_\theta\bk{\abs{\tilde X_m}^2} = \mc N A_{mm}(\theta),
\\
A(\theta) &\equiv V \SigmaX(\theta) V^\dagger,
\end{align}
where $\dagger$ denotes the conjugate transpose. To compute the
covariance matrix $E$, on the other hand, we use the law of total
variance and the Gaussian moment theorem \cite{mandel} to obtain
\begin{align}
E_{mn}(\theta) &= \delta_{mn} \expect_\theta(L_m) + \mc N^2 G_{mn}(\theta),
\\
G_{mn}(\theta) &\equiv |A_{mn}(\theta)|^2 + |B_{mn}(\theta)|^2,
\\
B(\theta) &\equiv V \SigmaX(\theta) V^\top.
\end{align}
These quantities can all be computed numerically.

If the input optical beam is squeezed, then the optimal measurement is
the unsqueezing of the output beam followed by SPC
\cite{noise_spec_pra}.  This case can be modeled by
multiplying the photon flux $\mc N$ by an enhancement factor due to
the squeezing and unsqueezing. As this modification is trivial
in theory, we focus on the coherent-state case hereafter.

In the SPLOT limit, $\{\tilde X_m: m \ge 0\}$ become independent
zero-mean Gaussian random variables and each $\tilde X_m$ has a
variance given by
\begin{align}
\expect_\theta\bk{|\tilde X_m|^2} &= S_X(\omega_m|\theta).
\end{align}
Since $X(t)$ is real, the positive-frequency components are related to
the negative-frequency components by
\begin{align}
\tilde X_m &= \tilde X_{-m}^*.
\end{align}
This relation implies that $L_m$ and $L_{-m}$ are not independent and
the likelihood function with respect to $\{L_m\}$ is difficult to
evaluate. To simplify the theory, consider instead a new
observation process given by
\begin{align}
N_m &\equiv \begin{cases} L_m + L_{-m}, & m > 0,\\ L_0, & m = 0\end{cases}
\end{align}
Now $\{N_m\}$ become independent random variables in the SPLOT limit
and are easier to analyze \cite{ng16}.  For $m > 0$, each $N_m$
observes the Bose--Einstein distribution
\begin{align}
P_{N_m}(n_m|\theta) &=  \frac{1}{\bar N_m(\theta)+1}\Bk{\frac{\bar N_m(\theta)}
{\bar N_m(\theta)+1}}^{n_m},
\\
\bar N_m(\theta) &\equiv 2 \mc N S_X(\omega_m|\theta).
\end{align}
$N_0 = L_0$, on the other hand, needs to be treated separately, since
$\tilde X_0$ is real. The probability distribution of $N_0$ turns out
to be \cite[Story~8.4.5]{blitzstein}
\begin{align}
P_{N_0}(n_0|\theta) &\propto 
\Bk{\frac{1}{1+2\bar N_0(\theta)}}^{1/2}
\Bk{\frac{2\bar N_0(\theta)}{1+2\bar N_0(\theta)}}^{n_0} ,
\\
\bar N_0(\theta) &\equiv \mc N S_X(0|\theta).
\end{align}
In the SPLOT limit, the negative log-likelihood function becomes
\begin{widetext}
\begin{align}
-\ln P_N^{(\mathrm{SPC,SPLOT})}(n|\theta) 
&= \frac{1}{2} \ln \Bk{1 + 2 \bar N_0(\theta)}
+ n_0 \ln \Bk{1 + \frac{1}{2\bar N_0(\theta)}}
+ \sum_{m> 0} \BK{\ln \Bk{1+\bar N_m(\theta)} + n_m \ln \Bk{1 + \frac{1}{\bar N_m(\theta)}}},
\label{loglike_SPC_SPLOT}
\\
\bar N_m(\theta) &= 
\begin{cases} 2 \mc N S_X(\omega_m|\theta), & m > 0,\\
\mc N S_X(0|\theta), & m = 0,
\end{cases}
\end{align}
while the Fisher information becomes \cite{ng16}
\begin{align}
J_{\mu\nu}^{(\mathrm{SPC,SPLOT})}(\theta)
&= \frac{T}{2\pi} 
\intall \frac{1}{2 + 1/[\mc N S_X(\omega|\theta)]}
\Bk{\pdv{}{\theta_\mu} \ln S_X(\omega|\theta)} 
\Bk{\pdv{}{\theta_\nu} \ln S_X(\omega|\theta)}
\dd\omega.
\label{JSPC_SPLOT}
\end{align}
\end{widetext}
Sec.~\ref{sec_fish} will compare this expression with the lower bound
given by Eq.~(\ref{stein}), while Sec.~\ref{sec_ML} will use
Eq.~(\ref{loglike_SPC_SPLOT}) to compute the maximum-likelihood
estimator for SPC given a frequency series
  $\{n_m: m \ge 0\}$. This SPLOT-based likelihood function is our
only option for computing the maximum-likelihood estimator for SPC,
since the exact likelihood function is intractable to our knowledge.

\subsection{Quantum limit}
Let the quantum state of a system at the initial time be $\ket{\phi}$
in some Hilbert space $\mc H$. Assume that the signal $X(t)$ perturbs
the system, such that the state at the final time conditioned on $X$
in the Schr\"odinger picture is given by
\begin{align}
\ket{\psi_X} &= U_M \dots U_1 \ket{\phi},
\\
U_j &= \exp\BK{-i\Bk{H_j- q_j X(t_j)}\delta t},
\label{U}
\end{align}
where $q_j$ and $H_j$ are self-adjoint operators on $\mc H$. We call
$\{q_j\}$ the generators. For example, if the system is an optical
beam under phase modulation, then $q_j$ is the photon-number operator
of the $j$th temporal mode, or if the system is a bosonic system
  under a force, then $q_j$ is a quadrature operator.  $H_j$, on the
other hand, is a Hamiltonian that models any additional dynamics of
the system.  The unconditional state at the final time becomes
\begin{align}
\rho(\theta) &= \expect_\theta\bk{\ket{\psi_X}\bra{\psi_X}}.
\label{rho}
\end{align}

We now use the extended-convexity property of the quantum Fisher
information to derive a quantum upper bound on the Fisher information
for any measurement, following Paper I and Ref.~\cite{alipour}.  Given
a mixed state of the form
\begin{align}
\rho(\theta) &= \expect_\theta\Bk{\sigma(W,\theta)} = 
\int \sigma(w,\theta) f_W(w|\theta) \dd^Mw,
\label{rho2}
\end{align}
where $f_W(w|\theta)$ is the probability density of a random variable
$W$ conditioned on $\theta$ and $\sigma(W,\theta)$ is a density
operator conditioned on $W$ and $\theta$, the extended-convexity bound
$J^{(\mathrm{EC})}$ on the quantum Fisher information
$J^{(\mathrm{Q})}$ in terms of $\rho(\theta)$ is given by
\cite{alipour,ng16}
\begin{align}
J^{(\mathrm{Q})}\Bk{\rho(\theta)} \le 
J^{(\mathrm{EC})}(\theta) &\equiv 
\expect_\theta\BK{J^{(\mathrm{Q})}\Bk{\sigma(W,\theta)} }
\nonumber\\&\quad
+ J\Bk{f_W(w|\theta)},
\label{JEC}
\end{align}
where $J^{(\mathrm{Q})}\Bk{\sigma(W,\theta)}$ is the quantum Fisher
information in terms of $\sigma(W,\theta)$,
$\expect_\theta\BK{J^{(\mathrm{Q})}\Bk{\sigma(W,\theta)}}$ is the
expectation in terms of $f_W$, and $J\Bk{f_W(w|\theta)}$ is the
classical Fisher information in terms of $f_W$. $J^{(\mathrm{Q})}$ and
therefore $J^{(\mathrm{EC})}$ are upper bounds on the Fisher
information for any measurement on a quantum system in state
$\rho(\theta)$. To obtain a tighter bound based on
$J^{(\mathrm{EC})}$, we exploit the fact that the same density
operator may admit many different ensembles
$(\sigma(w,\theta),f_W(w|\theta))$ in the form of Eq.~(\ref{rho2}),
each leading to a different $J^{(\mathrm{EC})}$. We then choose the
ensemble that gives the lowest $J^{(\mathrm{EC})}$.  Although it is
unknown in general how close the upper bound $J^{(\mathrm{EC})}$ is to
$J^{(\mathrm{Q})}$, our analysis will show that $J^{(\mathrm{EC})}$,
once optimized, turns out to be an excellent bound for our problem.

To consider a large class of ensembles for Eq.~(\ref{rho}) while
keeping $J^{(\mathrm{EC})}$ tractable, we transform $X$ to a new set
of zero-mean Gaussian random variables $W$ by
\begin{align}
X(t_j) &= \sum_m g_{jm}(\theta) W_m
\label{XW}
\end{align}
in terms of a real invertible matrix $g(\theta)$, leading to 
\begin{align}
\rho(\theta) &= 
\int \ket{\psi_{g(\theta) w}}\bra{\psi_{g(\theta) w}} f_W(w|\theta) \dd^M w.
\end{align}
$g(\theta)$ will be a variational parameter chosen to minimize
$J^{(\mathrm{EC})}$.  Let the covariance matrix of $W$ be
$\SigmaW(\theta)$. Then it is related to the covariance matrix
$\SigmaX(\theta)$ of $X$ by
\begin{align}
\SigmaX &= g \SigmaW g^\top,
&
\SigmaW &= g^{-1} \SigmaX g^{-\top}.
\end{align}
Focusing on the simple case of a scalar parameter $\theta$, the
extended-convexity bound becomes \cite{ng16}
\begin{align}
J^{(\mathrm{EC})} &= 4\expect_\theta\bk{ 
\langle\dot\psi_{gW}|\dot\psi_{gW}\rangle - \abs{\langle\psi_{gW}|\dot\psi_{gW}\rangle}^2} 
\nonumber\\&\quad
+ \frac{1}{2} \trace\bk{\SigmaW^{-1} \dot\SigmaW\SigmaW^{-1} \dot\SigmaW},
\end{align}
where the dot denotes $\pdv*{}{\theta}$. After some algebra, we find
\begin{align}
J^{(\mathrm{EC})} &= \trace
\bigg(h \SigmaX h^\top Q + h^2 + \SigmaX^{-1} h \SigmaX h^\top 
- 2 h^\top\gamma +\frac{\gamma^2}{2}\bigg),
\label{JEC2}
\end{align}
where 
\begin{align}
h &\equiv \dot g g^{-1},
\\
\gamma &\equiv \SigmaX^{-1} \dot\SigmaX,
\\
Q_{jk} &\equiv 4 \delta t^2 
\bk{\bra{\phi} \tilde q_j \circ \tilde q_k \ket{\phi}-\bra{\phi}\tilde q_j\ket{\phi}
\bra{\phi}\tilde q_k\ket{\phi}},
\label{Q}
\\
\tilde q_j &\equiv U_1^\dagger \dots U_j^\dagger q_j U_j \dots U_1.
\label{q_hei}
\end{align}
$h$ is related to the $g$ matrix in Eq.~(\ref{XW}), $\gamma$ is
related to the covariance matrix $\SigmaX$ of $X$, and $Q$ is
proportional to a discrete-time quantum covariance function of
the generators $\{\tilde q_j\}$ in the Heisenberg picture
conditioned on $X = gW$. In the derivation, $Q$ is assumed to be
independent of $W$. This is true when $\{q_j\}$ and $\{H_j\}$
  all commute, leading to $\tilde q_j = q_j$, or when $\{q_j\}$ are
  bosonic quadrature operators and $\{H_j\}$ are at most quadratic
  with respect to bosonic quadrature operators, such that
  $\{\tilde q_j\}$ remain quadratures and $Q$ does not depend on the
  displacement $X = g W$.

As $J^{(\mathrm{EC})}$ depends on the $g$ matrix through $h$, we now
choose an $h$ that minimizes $J^{(\mathrm{EC})}$ to obtain the
tightest upper bound. Taking the derivatives of $J^{(\mathrm{EC})}$
with respect to $h$ and setting them to zero, we find
\begin{align}
Q h \SigmaX  + h^\top + \SigmaX^{-1} h \SigmaX - \gamma = 0.
\label{opt_h}
\end{align}
This matrix equation can be solved numerically by vectorization
\cite[Chap.~4]{horn2};
let the solution be $\tilde h$. $J^{(\mathrm{EC})}$ becomes
\begin{align}
J^{(\mathrm{EC})} &= \trace\bk{\frac{\gamma^2}{2} - \tilde h^\top\gamma}.
\label{EC2}
\end{align}
It is straightforward to generalize this expression for multiple
parameters to obtain the information matrix
\begin{align}
J_{\mu\nu}^{(\mathrm{EC})} &= 
\frac{1}{2} \trace\bk{\gamma_\mu\gamma_\nu - \tilde h_\mu^\top\gamma_\nu
-\tilde h_\nu^\top\gamma_\mu},
\label{EC_matrix}
\\
\gamma_\mu &\equiv \SigmaX^{-1} \pdv{\SigmaX}{\theta_\mu},
\end{align}
where each $\tilde h_\mu$ satisfies
\begin{align}
Q \tilde h_\mu \SigmaX  + \tilde h_\mu^\top + \SigmaX^{-1} \tilde h_\mu \SigmaX - \gamma_\mu = 0.
\end{align}
We will not consider the multiparameter problem in this work however.

We can further simplify Eqs.~(\ref{opt_h}) and (\ref{EC2}) for the
phase-modulation problem, where $H_j = 0$ and $\tilde q_j = q_j$
  is the photon-number operator of the $j$th temporal mode. For
  a coherent initial state $\ket{\phi}$, the quantum covariance of
the photon-number operators is
\begin{align}
Q &= (4 \mc N\delta t) I,
\label{Q_coh}
\end{align}
where $I$ is the identity matrix. Assume that $\SigmaX$ is
proportional to the parameter $\theta$, such that
\begin{align}
\SigmaX &= \theta F
\end{align}
for a given positive-definite matrix $F$.  Assuming further that $h$
commutes with $F$ and $h = h^\top$, we obtain
\begin{align}
\gamma &= \frac{I}{\theta},
\\
\tilde h &= \frac{1}{\theta}\Bk{\bk{4 \mc N\theta\delta t } F + 2 I}^{-1},
\\
J^{(\mathrm{EC})} &= \frac{1}{2\theta^2} \trace\BK{I - 
\Bk{\bk{2 \mc N\theta\delta t } F + I}^{-1}}.
\label{EC}
\end{align}
Sec.~\ref{sec_fish} will use Eq.~(\ref{EC}) to compute the quantum
bound for finite times.

In the SPLOT limit, we can assume that $\SigmaX$, $\dot\SigmaX$, $g$,
$Q$, and $h$ are all circulant matrices, so that the
extended-convexity bound given by Eq.~(\ref{EC_matrix}) becomes
\cite{ng16}
\begin{align}
J^{(\mathrm{EC,SPLOT})}_{\mu\nu} 
&= \frac{T}{2\pi} \intall \frac{1}{2+1/(\mc N S_X)} 
\nonumber\\&\quad
\times \bk{\pdv{}{\theta_\mu} \ln S_X} \bk{\pdv{}{\theta_\nu} \ln S_X}\dd\omega.
\label{EC_SPLOT}
\end{align}
This expression is the same as Eq.~(\ref{JSPC_SPLOT}) for SPC in the
SPLOT limit. Since
  $J^{(\mathrm{SPC})} \le J^{(\mathrm{Q})} \le J^{(\mathrm{EC})}$,
  $J^{(\mathrm{SPC,SPLOT})} = J^{(\mathrm{EC,SPLOT})}$ implies that
  $J^{(\mathrm{SPC,SPLOT})} = J^{(\mathrm{Q,SPLOT})} =
  J^{(\mathrm{EC,SPLOT})}$, meaning that, in the SPLOT limit, SPC is
  quantum-optimal and the extended-convexity bound is perfectly
  tight.

\section{\label{sec_num}Numerical analysis}

\subsection{\label{sec_fish}Fisher information}
We now examine when the SPLOT-based Fisher-information expressions
become reliable at finite observation times by comparing them with the
finite-time results derived in Sec.~\ref{sec_theory}. For this
purpose, we model the signal $X(t)$ as an Ornstein–Uhlenbeck (OU)
process with power spectral density
\begin{align}
S_X(\omega|\theta) &= \frac{2\theta}{\omega^2 + 1},
&
\theta &= \expect_\theta\Bk{X(t)^2},
\end{align}
such that $\theta$ is the variance of $X$. All the variables,
including $\theta$, the observation time $T$, and the frequency
variable $\omega$, are normalized with respect to the decay rate of
the process. Defining a signal-to-noise-ratio quantity $C$ in terms of
$\theta$ and the average photon flux $\mc N$ as
\begin{align}
C &\equiv 8\theta \mc N
\label{C}
\end{align}
and using Eqs.~(\ref{Jhom_SPLOT}), (\ref{JSPC_SPLOT}), and
(\ref{EC_SPLOT}), Paper I has found that, in the SPLOT limit,
\begin{align}
J^{(\mathrm{hom,SPLOT})} &= \frac{T}{8\theta^2} \frac{C^2}{(1+C)^{3/2}},
\label{Jhom_SPLOT2}
\\
J^{(\mathrm{SPC,SPLOT})} &= J^{(\mathrm{EC,SPLOT})}
= \frac{T}{8\theta^2} \frac{C}{\sqrt{1+C/2}}.
\label{JSPC_SPLOT2}
\end{align}
Define the weak-signal regime as $C \ll 1$.  In this regime,
$J^{(\mathrm{SPC,SPLOT})} \approx J^{(\mathrm{hom,SPLOT})}/C$,
suggesting a substantial enhancement by SPC. To confirm this
prediction, we compare Eqs.~(\ref{Jhom_SPLOT2}) and
(\ref{JSPC_SPLOT2}) with the finite-time expressions given by
Eq.~(\ref{Jhom}), Eq.~(\ref{stein}), and Eq.~(\ref{EC}) at
$\theta = 1$; the Fisher information at other values of $\theta$ can
be obtained by a rescaling argument as long as $C$ is the same, as
shown in Appendix~\ref{sec_rescale}. We perform the numerical analysis
using the MATLAB software on a personal computer.  The discrete time
interval $\delta t$ is taken to be $0.01$ throughout the analysis.
The results are shown in Fig.~\ref{fisher_plots}.

\fig{}{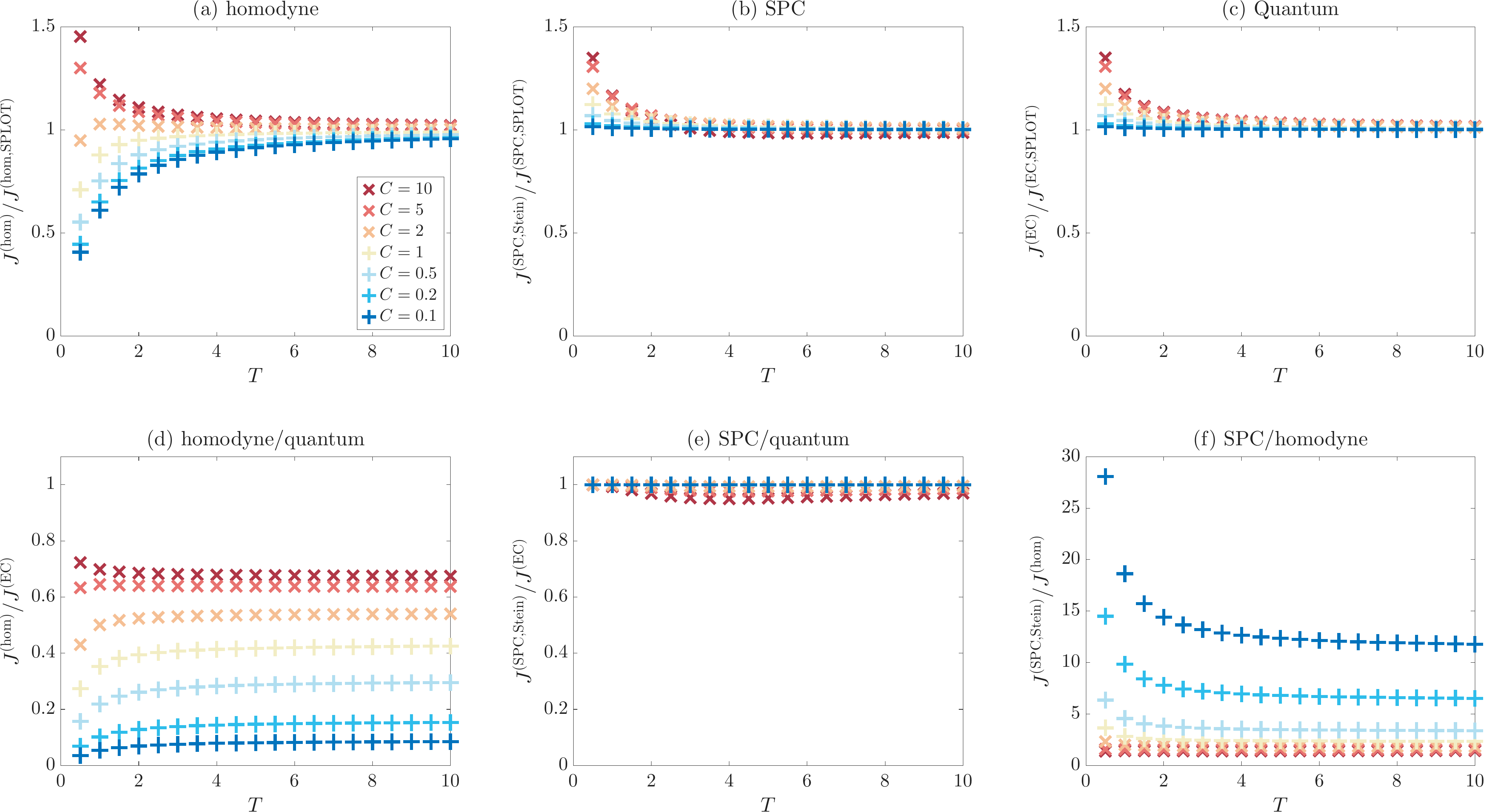}{(a) The ratio of the homodyne information
  $J^{(\mathrm{hom})}$ given by Eq.~(\ref{Jhom}) to the SPLOT limit
  $J^{(\mathrm{hom,SPLOT})}$ given by Eq.~(\ref{Jhom_SPLOT2}).  (b)
  The ratio of the lower bound $J^{(\mathrm{SPC,Stein})}$ given by
  Eq.~(\ref{stein}) for spectral photon counting (SPC) to the SPC
  information in the SPLOT limit $J^{(\mathrm{SPC,SPLOT})}$ given by
  Eq.~(\ref{JSPC_SPLOT2}).  (c) The ratio of the extended-convexity
  upper bound $J^{(\mathrm{EC})}$ given by Eq.~(\ref{EC}) to the SPLOT
  limit $J^{(\mathrm{EC,SPLOT})}$ given by Eq.~(\ref{JSPC_SPLOT2}).
  (d) The ratio of $J^{(\mathrm{hom})}$ to $J^{(\mathrm{EC})}$,
  demonstrating the suboptimality of homodyne detection for low $C$.
  (e) The ratio of $J^{(\mathrm{SPC,Stein})}$ to $J^{(\mathrm{EC})}$,
  demonstrating that the two bounds must be tight if they are so
  close. (f) The ratio of $J^{(\mathrm{SPC,Stein})}$ to
  $J^{(\mathrm{hom})}$, demonstrating the superiority of SPC over
  homodyne detection.  The different plots in each figure are for
  different values of a signal-to-noise-ratio quantity $C$ defined by
  Eq.~(\ref{C}).  All quantities, such as the observation time $T$,
  are normalized with respect to the signal decay rate.}

Figs.~\ref{fisher_plots}(a)--(c) compare the finite-time Fisher
information quantities with their SPLOT limits as a function of the
observation time $T$. While there are appreciable differences for
short times, all curves converge smoothly to their SPLOT limits as $T$
increases.  These plots show that the SPLOT-based Fisher-information
expressions are simple and adequate approximations of the exact
quantities, once the observation time $T$ (normalized with respect to
the decay rate of $X$) exceeds roughly $5$.

Figs.~\ref{fisher_plots}(d)--(f) further compare the finite-time
Fisher-information quantities for homodyne detection, SPC, and the
quantum bound without the SPLOT
approximation. Fig.~\ref{fisher_plots}(d) plots the ratio of the
homodyne information $J^{(\mathrm{hom})}$ given by Eq.~(\ref{Jhom}) to
the quantum upper bound $J^{(\mathrm{EC})}$ given by Eq.~(\ref{EC}),
showing that homodyne detection is suboptimal with respect to the
quantum limit, especially for low $C$.  Fig.~\ref{fisher_plots}(e)
plots the ratio of the lower bound $J^{(\mathrm{SPC,Stein})}$ for SPC
given by Eq.~(\ref{stein}) to the quantum upper bound
$J^{(\mathrm{EC})}$. Since the exact SPC information
  $J^{(\mathrm{SPC})}$ obeys
  $J^{(\mathrm{SPC,Stein})} \le J^{(\mathrm{SPC})} \le
  J^{(\mathrm{Q})} \le J^{(\mathrm{EC})}$, the closeness of
  $J^{(\mathrm{SPC,Stein})}$ and $J^{(\mathrm{EC})}$ shown in
  Fig.~\ref{fisher_plots}(e) implies that $J^{(\mathrm{SPC,Stein})}$
  must be a tight lower bound on $J^{(\mathrm{SPC})}$,
  $J^{(\mathrm{EC})}$ must be a tight upper bound on the quantum
  Fisher information $J^{(\mathrm{Q})}$, and $J^{(\mathrm{SPC})}$ must
  be close to $J^{(\mathrm{Q})}$.  Finally,
Fig.~\ref{fisher_plots}(f) plots the ratio of the SPC bound
$J^{(\mathrm{SPC,Stein})}$ to the homodyne information
$J^{(\mathrm{hom})}$, demonstrating that the superiority of SPC over
homodyne detection remains significant for low $C$ and short $T$.

\subsection{\label{sec_ML}Maximum-likelihood estimation}
To demonstrate that the Fisher information and the resulting
Cram\'er-Rao bounds are reliable benchmarks for the estimation
performances in finite time, we generate many samples of the
observation processes for both homodyne detection and SPC and compute
the maximum-likelihood estimator for each sample, following the
same setting as that in Sec.~\ref{sec_fish}. We again assume
$\theta = 1$; the errors for other values of $\theta$ can be obtained
by a rescaling argument as long as $C$ is the same, as per
Appendix~\ref{sec_rescale}.  We use the SPLOT-based likelihood
functions given by Eqs.~(\ref{loglike_hom_SPLOT}) and
(\ref{loglike_SPC_SPLOT}), so that the estimators can be computed
within a reasonable timeframe for long observation times and a large
number of samples. By performing the parameter estimation for a large
number of samples and averaging the errors, we can estimate the
expected mean-square error (MSE), given by
\begin{align}
\MSE(\theta) &\equiv \expect_\theta\Bk{\bk{\check\theta - \theta}^2},
\label{MSE}
\end{align}
where $\check\theta$ is the estimator as a function of the time
  series $\{Y(t)\}$ for homodyne detection or the frequency series
  $\{N_m\}$ for SPC.

Fig.~\ref{spectroscopy_MLE3} plots the results for four different
values of $C$. All plots are in log-log scale. Note that the
observation times here are now much longer than those considered in
Fig.~\ref{fisher_plots}.  For each observation time $T$, the
number of generated samples of an observation process is chosen to
maintain a satisfactory precision for the estimates. To quantify this
precision, we compute 95\% bootstrap confidence intervals using the
\texttt{bootci} function on MATLAB with $10,000$ bootstrap
replications;
the resulting confidence intervals are shown as error bars in
Fig.~\ref{spectroscopy_MLE3}.


\fig{}{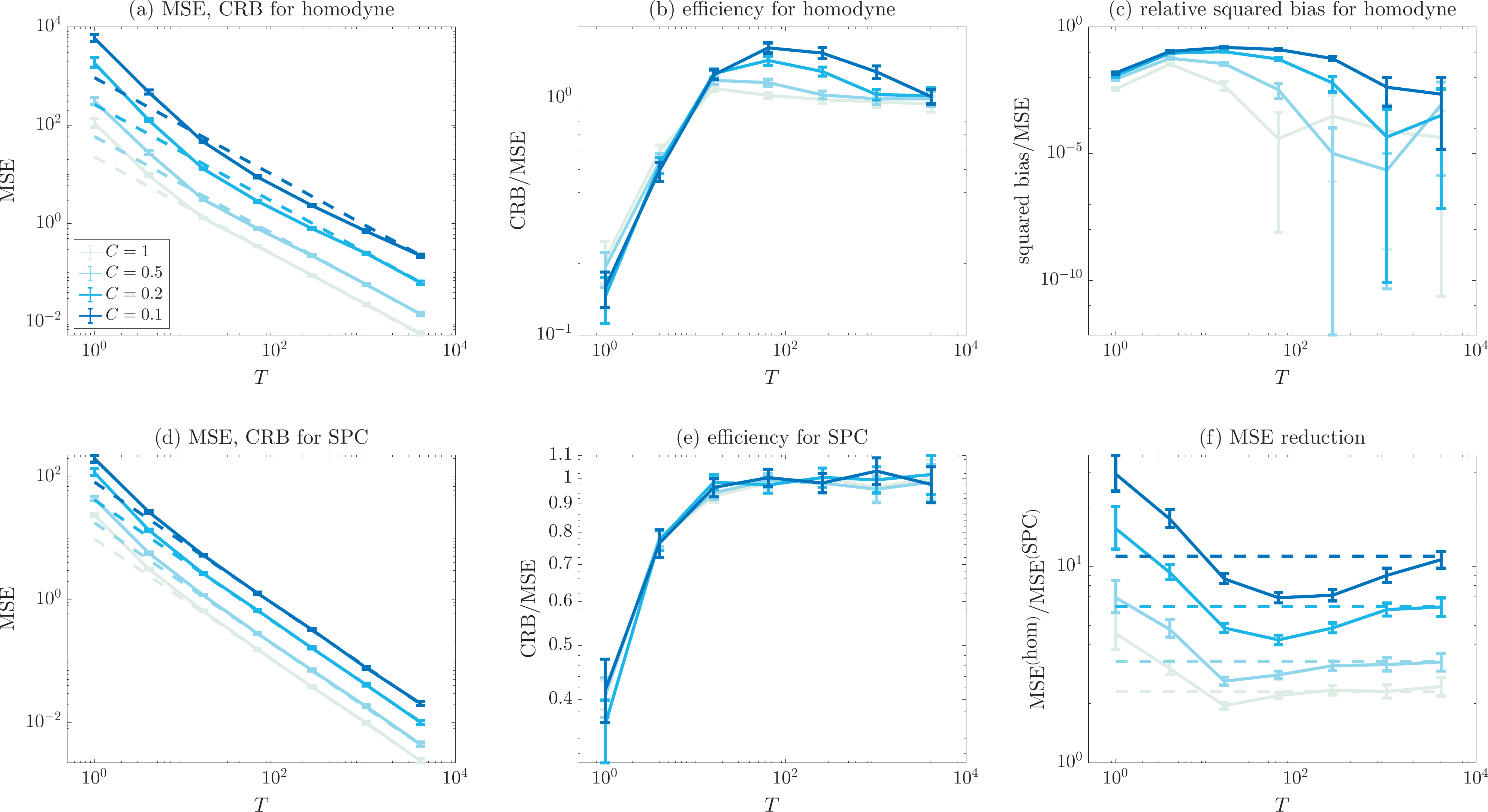}{(a) The homodyne mean-square error (MSE)
  computed by Monte-Carlo simulations. The dash curves plot the
  SPLOT-based Cram\'er-Rao bound (CRB) $1/J^{(\mathrm{hom,SPLOT})}$
  computed from Eq.~(\ref{Jhom_SPLOT2}). (b) The homodyne efficiency,
  defined as the ratio of the SPLOT-based CRB
  $1/J^{(\mathrm{hom,SPLOT})}$ to the homodyne MSE. (c) The ratio of
  the squared bias to the MSE for homodyne detection. (d) and (e) are
  for SPC and follow the same formats as those of (a) and (b).  (f)
  The ratio of the homodyne MSE to the SPC MSE, demonstrating the
  superiority of SPC for the weak-signal regime (low $C$) and finite
  observation time $T$. The dash curves plot the ratio
  $J^{(\mathrm{SPC,SPLOT})}/J^{(\mathrm{hom,SPLOT})}$. All plots are
  against the total observation time $T$ and in log-log scale. The
  straight lines connecting the data points are guides for eyes. Each
  error bar indicates the 95\% bootstrap confidence interval computed
  from $10,000$ bootstrap replications.}

Figs.~\ref{spectroscopy_MLE3}(a) and (d) compare the numerical errors
of the maximum-likelihood estimators with the Cram\'er-Rao bounds for
homodyne detection and SPC, respectively. The dashed curves in
Figs.~\ref{spectroscopy_MLE3}(a) and (d) are the SPLOT-based
Cram\'er-Rao bounds given by $1/J^{(\mathrm{hom,SPLOT})}$ and
$1/J^{(\mathrm{SPC,SPLOT})} = 1/J^{(\mathrm{EC,SPLOT})}$,
respectively.  The errors all approach the bounds for increasing $T$.
This asymptotic behavior is expected, since the maximum-likelihood
estimator is asymptotically unbiased and efficient in the SPLOT limit
\cite{shumway_stoffer}. A key observation is that the trends of the
errors are all smooth and the differences between the errors and the
bounds are never substantial, unlike the large discrepancies that may
occur for other problems \cite{bell,qzzb}.  To further demonstrate the
convergence, we plot the efficiency, defined as the ratio of the
SPLOT-based bound to the error, for various $C$ and $T$ in
Figs.~\ref{spectroscopy_MLE3}(b) and (e). They all converge smoothly to
$100\%$ for increasing $T$.

For a low $C$ and an intermediate $T$, Figs.~\ref{spectroscopy_MLE3}(a)
and (b) show that the homodyne error may violate the Cram\'er-Rao
bound slightly and the efficiency may go above $100\%$. This behavior
is caused by the bias of the estimator, defined as
\begin{align}
\mathrm{bias}(\theta) &\equiv \expect_\theta\bk{\check\theta} - \theta.
\end{align}
Fig.~\ref{spectroscopy_MLE3}(c) plots the squared bias relative to the
error for homodyne detection. Notice that, for the $C$ and $T$ values
at which the error violates the bound, the bias is relatively
high. This bias allows the estimator to violate the bound, although
the error eventually returns to the bound for a longer $T$, as it
should.

Last but not the least, Fig.~\ref{spectroscopy_MLE3}(f) compares the
homodyne and SPC performances by plotting the ratio of the homodyne
error to the SPC error. This result demonstrates convincingly that the
SPC remains superior to homodyne detection for low $C$ and finite
times. The dashed curves in Fig.~\ref{spectroscopy_MLE3}(d) correspond
to the ratio
\begin{align}
\frac{J^{(\mathrm{SPC,SPLOT})}}{J^{(\mathrm{hom,SPLOT})}}
= \frac{J^{(\mathrm{EC,SPLOT})}}{J^{(\mathrm{hom,SPLOT})}}.
\end{align}
Its closeness to the actual error ratio shows that the SPLOT-based
Fisher-information expressions are adequate benchmarks for the
estimation performances.

\section{Conclusion}
In this work, we have numerically verified that the Fisher information
and the Cram\'er-Rao bounds derived in Paper I using the SPLOT
approximation are adequate benchmarks for the estimation errors in
finite time. Most importantly, the results demonstrate that SPC
remains superior to homodyne detection for finite time in the
weak-signal regime. Although the experimental challenges are
significant \cite{mcculler22,vermeulen25}, we can now be more
confident that, once the technical challenges are overcome, our
predictions based on the SPLOT approximation will hold in reality.

While we have assumed very simple models for the signal and the noise
in our numerical study, the general formalism presented in this work
should allow our methods to be extended for more complicated Gaussian
states, stationay Gaussian signal processes, and multiple
parameters. If, on the other hand, the state is non-Gaussian or the
signal is nonstationary or non-Gaussian, then the quantum limits and
the optimal measurements are outside the scope of our current
formalism and await further studies; other researchers have made some
recent progress on that front \cite{gardner25,gardner26}.

\appendix
\section{\label{sec_rescale}Rescaling}
For the Gaussian process $Y$ defined in Sec.~\ref{sec_hom}, its
covariance matrix is
\begin{align}
\Sigma(\theta,\mc N) &= \theta F + \frac{1}{4\mc N\delta t} I,
\end{align}
where $F$ is the covariance matrix of $X$ when $\theta = 1$ and
$\mc N$ is the average photon flux of a continuous-wave coherent
state. Let the probability density of $Y$ be $f_Y(y|\theta,\mc
N)$. Define the maximum-likelihood estimator as
\begin{align}
\check\theta(y,\mc N) &\equiv \argmax_\theta f_Y(y|\theta,\mc N),
\label{MLE}
\end{align}
and its mean-square error (MSE) as
\begin{align}
\MSE^{(\mathrm{hom})}(\theta,\mc N) &\equiv 
\int \Bk{\check\theta(y,\mc N)-\theta}^2 f_Y(y|\theta,\mc N)\dd^M y.
\end{align}
Now define a new process
\begin{align}
\eta(t) &\equiv \frac{1}{\sqrt{c}} Y(t)
\end{align}
for some constant $c > 0$. Then its covariance matrix
$\tilde\Sigma_{jk} \equiv \expect_\theta[\eta(t_j)\eta(t_k)]$ can be
expressed as
\begin{align}
\tilde\Sigma &= \frac{1}{c} \SigmaX + \frac{1}{c}\SigmaZ
= \frac{\theta}{c} F + \frac{1}{4 c \mc N \delta t} I
= \Sigma(\theta/c,c\mc N),
\end{align}
which is the same as the original $\Sigma$ but with rescaled
parameters $(\theta/c,c\mc N)$. There are now two ways of writing the
probability density $f_{\eta}(u|\theta)$:
\begin{enumerate}
\item Use the change-of-variable formula to write
\begin{align}
f_{\eta}(u|\theta) &= c^{M/2} f_Y(\sqrt{c}u|\theta,\mc N).
\label{change_var}
\end{align}
\item Regard $\eta$ as $Y$ with rescaled parameters to write
\begin{align}
f_{\eta}(u|\theta) &= f_Y(u|\theta/c,c\mc N).
\label{rescaled}
\end{align}
\end{enumerate}
It follows that there are also two ways of writing the
maximum-likelihood estimator $\check\theta(u)$ and the MSE given
$\eta = u$:
\begin{enumerate}
\item Following Eqs.~(\ref{MLE}) and (\ref{change_var}), we obtain
\begin{align}
\check\theta(u) &\equiv \argmax_\theta f_{\eta}(u|\theta) 
\\
&= \argmax_\theta c^{M/2} f_Y(\sqrt{c}u|\theta,\mc N) \\
&=\check\theta(\sqrt{c}u,\mc N).
\end{align}
The error becomes
\begin{align}
\MSE(\theta) &\equiv \int \Bk{\check\theta(u) -\theta}^2 f_{\eta}(u|\theta) \dd^Mu
\\
&= \int \Bk{\check\theta(\sqrt{c}u,\mc N) - \theta}^2 
c^{M/2} f_Y(\sqrt{c}u|\theta,\mc N) \dd^M u
\\
&= \int \Bk{\check\theta(y,\mc N) - \theta}^2 
f_Y(y|\theta,\mc N) \dd^M y
\\
&= \MSE^{(\mathrm{hom})}(\theta,\mc N).
\label{error1}
\end{align}
This equality makes sense, as we have simply multiplied the original
process $Y$ by a constant.

\item Following Eqs.~(\ref{MLE}) and (\ref{rescaled}), we obtain
\begin{align}
\check\theta(u) &\equiv \argmax_\theta f_{\eta}(u|\theta) 
\\
&= \argmax_\theta f_Y(u|\theta/c,c\mc N)
\\
&=c \argmax_{\phi} f_Y(u|\phi,c\mc N)
\\
&= c\check\theta(u,c \mc N).
\end{align}
The error becomes
\begin{align}
\MSE(\theta) &\equiv \int \Bk{\check\theta(u) -\theta}^2 f_{\eta}(u|\theta) \dd^Mu
\\
&= \int \Bk{c\check\theta(u,c \mc N) -\theta}^2 f_Y(u|\theta/c,c\mc N) \dd^Mu
\\
&= c^2\MSE^{(\mathrm{hom})}(\theta/c,c\mc N).
\label{error2}
\end{align}
This equality shows that $\MSE(\theta)$ is equal to the error at a
different set of parameters $(\theta/c,c\mc N)$.
\end{enumerate}
Combining Eqs.~(\ref{error1}) and (\ref{error2}), we obtain
\begin{align}
\MSE^{(\mathrm{hom})}(\theta,\mc N) &= c^2\MSE^{(\mathrm{hom})}(\theta/c,c\mc N).
\end{align}
In particular, if $c = \theta$,
\begin{align}
\MSE^{(\mathrm{hom})}(\theta,\mc N) &= \theta^2 \MSE^{(\mathrm{hom})}(1,\theta\mc N).
\label{MSE_rescaled}
\end{align}
This formula shows that we can evaluate the maximum-likelihood
estimator and its error at $\theta = 1$ and a rescaled photon-flux
parameter $\theta \mc N$; the error at any other $\theta$ and the same
$\theta \mc N$ value can be obtained from Eq.~(\ref{MSE_rescaled}).
The same argument holds if we use the SPLOT approximation of the
likelihood function to evaluate the estimator.

Using Eq.~(\ref{rescaled}), we find that the Fisher information
follows a similar scaling law:
\begin{align}
J^{(\mathrm{hom})}(\theta,\mc N) &= \frac{1}{c^2} J^{(\mathrm{hom})}(\theta/c,c\mc N)
\\
&= \frac{1}{\theta^2} J^{(\mathrm{hom})}(1,\theta\mc N).
\end{align}

The argument for SPC is similar. Let the probability distribution of
the SPC outcome be $P_L(l|\theta,\mc N)$.  Following
Sec.~\ref{sec_SPC}, we find that
\begin{align}
P_L(l|\theta,\mc N) &= P_L(l|\theta/c,c\mc N).
\end{align}
Then we can use the same argument as above to obtain
\begin{align}
\MSE^{(\mathrm{SPC})}(\theta,\mc N) &= c^2\MSE^{(\mathrm{SPC})}(\theta/c,c\mc N) \\
&= \theta^2\MSE^{(\mathrm{SPC})}(1,\theta\mc N),
\\
J^{(\mathrm{SPC})}(\theta,\mc N) &= \frac{1}{c^2} J^{(\mathrm{SPC})}(\theta/c,c\mc N)
\\
&= \frac{1}{\theta^2} J^{(\mathrm{SPC})}(1,\theta\mc N).
\end{align}

\bibliography{research4}

\end{document}